\documentclass{tlp}
\usepackage{epsf}
\usepackage{url} 

\begin{document} 
\bibliographystyle{tlp}
\newcommand{\FIG}[4]{
\begin{figure}[htbp]
\begin{center}
\leavevmode
{ \setlength{\epsfysize}{#2 in} \setlength{\epsfxsize}{#3 in} 
  \epsffile{#1.eps}
}
\end{center}
\caption{#4}\label{#1}
\end{figure}
}


\newcommand{\seq}[2]{{#1} \vdash {#2}}
\newcommand{\sseq}[3]{{#1};{#2} \vdash {#3}}
\newcommand{\regle}[3]{
\begin{tabular}[b]{c}
#1\\
\hline
#2\\
\end{tabular}\,\,\raisebox{1.5ex}{$#3$}
}

\newcommand{\split}[4]{
\begin{tabular}[b]{c}
\begin{tabular}[b]{c}
#1
\end{tabular}
\begin{tabular}[b]{c}
#2
\end{tabular}\\
\hline
#3\\
\end{tabular}\,\,\raisebox{1.5ex}{$#4$}
}

\newcommand{\Regle}[3]{
#1
$\overline{\mbox{#2}}^{\,#3}$\\
}

\newcommand{\FirstRegle}[3]{
\begin{tabular}[b]{c}
#1
$\overline{\mbox{#2}}^{\, #3}$\\
\end{tabular}
}

\newcommand{\FirstSplit}[4]{
\begin{tabular}[b]{c}
\begin{tabular}[b]{c}
#1
\end{tabular}
\begin{tabular}[b]{c}
#2
\end{tabular}\\
$\overline{\mbox{#3}}^{\, #4}$\\
\end{tabular}
}

\newcommand{\Split}[4]{
\begin{tabular}[b]{c}
#1
\end{tabular}
\begin{tabular}[b]{c}
#2
\end{tabular}\\
$\overline{\mbox{#3}}^{\, #4}$\\
}

\def\comment#1{}

\newcommand{\mf}[1]{$#1$}

\title[High-Level Networking with Mobile Code and First Order Continuations]{High-Level Networking with Mobile Code and First Order AND-Continuations}
\author[Paul Tarau and Veronica Dahl]
{PAUL TARAU \\
Department of Computer Science\\
University of North Texas\\
P.O. Box 311366\\
Denton, Texas 76203\\
E-mail: {\em tarau@cs.unt.edu}\\
\and
VERONICA DAHL \\
Logic and Functional Programming Group\\
Department of Computing Sciences\\
Simon Fraser University\\
E-mail: {\em veronica@cs.sfu.ca}\\
}

\maketitle{}

\begin{abstract}
We describe a scheme for moving living code between a set
of distributed processes coordinated with
unification based Linda operations, and its
application to building a comprehensive Logic programming based
Internet programming framework.
{\em Mobile threads} are implemented by capturing
first order continuations
in a compact data structure sent over the network.
Code is fetched lazily from its original base turned into a server
as the continuation executes at the remote site.
Our code migration techniques,
in combination with a dynamic recompilation scheme,
ensure that heavily used code moves up smoothly on a speed
hierarchy while volatile dynamic code is kept in a 
quickly updatable form.
Among the examples, we describe how to build programmable client
and server components (Web servers, in particular)
and mobile agents.

\end{abstract}

\section{Introduction} \label{intro}

{\em Data mobility} has been present since
the beginning of networked computing, and is
now used in numerous applications --
from remote consultation of a database, to
Web browsing.

{\em Code mobility} followed, often made transparent
to users as with network files systems (i.e. Sun's NFS).
Java's ability to execute applets directly in client browsers,
can be seen as its most recent incarnation.

Migrating the state of the computation from one machine or process to
another, still requires a separate set of tools. Java's remote method
invocations (RMI) add {\em control mobility}
and a (partially) automated form of {\em object mobility}
i.e. integrated code (class) and data (state) mobility.
The Oz 2.0 distributed programming proposal of \cite{DOZmobility}
makes {\em object mobility} more
transparent, although the mobile entity is still
the state of the objects, not ``live'' code.

Mobility of ``live code'' is called {\em computation mobility} 
\cite{cardelli97:mobile}.
It requires interrupting the execution, moving the state of a runtime system
(stacks, for instance) from one site to another, and then resuming
execution. Clearly, for some languages, this can be hard or completely
impossible to achieve.

Telescript and General Magic's
Odyssey \cite{odissey} agent programming
framework, IBM's Java based {\em aglets}, as well as
Luca Cardelli's Oblique \cite{migratory},
have pioneered implementation technologies with
{\em computation mobility}.

This paper will show that we can achieve
full {\em computation mobility} through our
{\em mobile threads}, with no need for a specially designed new language.
It is implemented by a surprisingly small, source level
modification of the BinProlog system, which takes advantage of
the availability of {\em first order
continuations}\footnote{I.e. continuations (representations 
of future computations)
accessible as an ordinary data structure - a 
Prolog term in this case.} 
as well as
of BinProlog's high level networking primitives.
Mobile threads complete our Logic Programming based Internet
programming infrastructure, built in view of
creating Prolog components which can interoperate
with mainstream languages and programming environments.
{\em Mobile threads} can be seen as a refinement of {\em mobile
computations}, as corresponding to {\em mobile partial computations}
of any granularity. {\em Mobile agents} can be seen as a collection
of synchronized {\em mobile threads} sharing common
state \cite{TD96:coord}. We achieve synchronization using
a variant of the Linda coordination protocol.

The paper is organized as follows:
\begin{itemize}
\item section \ref{infra} describes our networking infrastructure 
and Linda based client/server components
\item Section \ref{mobcode} describes our code migration and
  code acceleration techniques (dynamic recompilation)
\item Section \ref{mobcomp} describes our mobile computation mechanism, as follows: 
   subsection \ref{engines} introduces engines and threads,
   subsection \ref{bin}) reviews the underlying binarization mechanism
     used to implement our first order continuations,
   subsection \ref{mobthreads} explains how we implement thread mobility
   by capturing continuations (subsection \ref{capt})
   and moving them from their base to their target
     (subsection \ref{contmoving}), how this can be emulated
     with remote predicate calls (subsection \ref{mobemu}) and how mobile 
     agents can be built within our framework (subsection \ref{mobags})
\item section \ref{rel} discusses related work
\item section \ref{conc} presents our conclusions and future work
\end{itemize}

The main ``paradigm independent'' novelties of our contribution are:

\begin{itemize}
\item use of first order continuations for implementing {\em mobile computations}
\item a flexible thread mobility algorithm expressed in terms of client-server 
 role alternation and communication through Linda operations
 \item a technique, based on intuitionistic assumptions, for dealing with
 complex networking code component-wise
\end{itemize}

Our contributions are synergetically integrated into a powerful agent building
infrastructure, which brings together logic programming
based knowledge processing, Linda-style coordination, 
and live code migration through mobile threads.

\section{Basic Linda and Remote Execution Operations} \label{infra}

\subsection{Coordination of Linda clients}

Our networking constructs are built on top of the popular
Linda \cite{linda89,Cia94b,dbt95a,Cia96,JavaSpaces} coordination framework,
enhanced with unification based pattern
matching, remote execution and a set of simple
client-server components
merged together into a scalable peer-to-peer
layer, forming a `web of interconnected worlds':

{\small \begin{verbatim}
out(X): puts X on the server
in(X):  waits until it can take an object 
        matching X from the server
all(X,Xs): reads the list Xs matching X 
        currently on the server
        
\end{verbatim}}

\FIG{linda}{2.25}{3.5}{Basic Linda operations}

{\flushleft The} presence of the {\tt all/2} collector
avoids the need for backtracking over multiple
remote answers.
Note that the only blocking operation is {\tt in/1}.
Typically, distributed programming with Linda coordination
follows consumer-producer patterns (see Fig. \ref{linda}) 
with added flexibility over message-passing communication
through associative search.
Blocking {\tt rd/1}, which waits until a matching term becomes
available, without removing it, is easily
emulated in terms of {\tt in/1} and {\tt out/1},
while non-blocking {\tt rd/1} is emulated with {\tt all/2}.

\subsection{Remote Execution Mechanisms}

The implementation of arbitrary remote execution is easy
in a Linda + Prolog system, due to Prolog's {\em metaprogramming}
abilities, which allow us to send arbitrary Prolog terms over the
network in a uniform way, without the need for implementing 
complex serialization/remote object mechanisms.
Our primitive remote call operation is:

{\small \begin{verbatim}
host(Other_machine)=>>
   remote_run(Answer,RemoteGoal).
\end{verbatim}}

\noindent It implements deterministic {\em remote predicate calls}
with (first)-answer or {\tt `no'} returned to the calling site.

For instance, to iterate over the set of servers forming
the receiving end of our `Web of Worlds', after retrieving
the list from a `master server' which constantly monitors
them making sure that the list reflects login/logout
information, we simply override {\tt host/1} and {\tt port/1} with
intuitionistic implication {\tt =>>} \cite{bp7advanced,DT97:AGNL}:

{\small \begin{verbatim}
ask_all_servers(Channel,Servers,Query):-
  member(server_id(Channel,H,P),Servers),
  host(H)=>>port(P)=>>
    ask_a_server(Query,_),
  fail;true.
\end{verbatim}}

\noindent Note that a {\tt Channel} pattern is used to select a subset
of relevant servers, and in particular, when {\tt Channel} is
a ``match all'' free logical variable,  all of them.
By using term subsumption this allows building sophisticated "publish/subscribe"
communication pattern hierarchies.

\subsection{Servants: basic Linda agents} \label{servant}

\noindent A {\em servant} is one of the simplest possible {\em agents},
which pulls commands from a server and runs them locally:

{\small \begin{verbatim}
servant:-
  in(todo(Task)),
  call(Task),
  servant.
\end{verbatim}}

\noindent Note that {\em servant} is  started as a background thread.
No `busy wait' is involved,
as the servant's thread
blocks until {\tt in(todo(Task))}
succeeds.
More generally, distributed event processing is implemented
by creating a `watching' agent attached to a thread
for each pattern.

As {\em servants}
pulling commands are operationally indistinguishable from
{\em servers} acting upon clients' requests, they
can be used as emulators for {\em servers}.
A class of obvious applications of this ability
is their use as pseudo-servers running on
machines with dynamically allocated
IP addresses (as offered by most ISP today),
laying behind firewalls.
This mechanism also works when, because of
security restrictions, server components
cannot be reached from outside, as in the case of
{\em Java applets} which cannot listen on ports of
the client side machine.

\subsection{Server side code}

Servants as well as other clients can connect to
BinProlog {\em servers}.
Higher order {\em call/N} \cite{mycroft:poli}, 
combined with intuitionistic assumptions {\tt ``=>>''},
are used to pass arbitrary {\em interactors} to
generic server code:

{\small \begin{verbatim}

run_server(Port):-
  new_server(Port,Server),
  register_server(Port),
  server(Server)=>>server_loop,
  close_socket(Server).

server_loop:-
  repeat,
    interact,   
  assumed(server_done),
  !.

interact:-
  assumed(Interactor),
  assumed(Server),
  % higher-order call to interactor 
  call(Interactor,Server).
\end{verbatim}}

{\flushleft Note} the use of a specialized {\em server-side} interpreter
{\tt server\_loop}, configurable through the use of higher-order 
`question/answer' closures we have called {\em interactors}.

The components of a `generic' default server can be 
overridden through the use of {\em intuitionistic
implication} to obtain customized special purpose servers.
The use of intuitionistic implications  
 (pioneered by
Miller's work \cite{Miller89Lex}) helps to overcome 
(to some extent) Prolog's lack
of object oriented programming facilities,
by allowing us to `inject' the right {\em interactor} into
the generic (and therefore reusable) interpreter.
BinProlog's {\tt ``=>>''} temporarily assumes a clause
in {\tt asserta} order, i.e. at the beginning of the predicate.
The assumption is scoped to be only usable to prove its
right side goal and vanishes on backtracking.
We refer to \cite{bp7advanced,TDF:asian96,DT97:AGNL}
for more information on assumptions and their applications.

\subsection{Modular HTTP server component building: {\em do or delegate}}

We will show in this section a typical application of our component based server
building technology: how to enhance efficiently the HTTP protocol, to handle
server side Prolog scripts directly, without using CGIs or vendor specific
server side extensions.

The top goal of the HTTP server looks as follows:

{\small \begin{verbatim}
run_http_server:-
  server_action(http_server_action)=>>run_server.

http_server_action(ServiceSocket):-
  socket(ServiceSocket)=>>http_server_step(ServiceSocket).
\end{verbatim}}

The {\tt http\_server\_action} is passed to the inner server
loop using intuitionistic implication. This allows reusing
general server logic, independently of a particular protocol.
The action itself is described in {\tt http\_server\_step}: it
consists of preparing a fall-back mechanism to a standard
HTTP server, unless the request is for a file recognized as
Prolog code, using the {\em redirection} facilities built
in the HTTP protocol. 

{\small \begin{verbatim}
http_server_step(Socket):-
  ( assumed(fallback_server(FallBackServer))->true
  ; FallBackServer="http://localhost:80"
  ),
  server_try(Socket,sock_readln(Socket,Question)),
  http_get_client_header(Socket,Header),
  http_process_query(Socket,Question,Header,FallBackServer).
\end{verbatim}}

Our very simple query processor uses Assumption Grammars \cite{DT97:AGNL}.
Their ability to handle multiple DCG streams \cite{TDF:asian96} is instrumental,
as we use more than one independent grammar processor in the process.

{\small \begin{verbatim}
http_process_query(Socket,Qs,Css,FallBackServer):-
  #<Qs, % sets input string from grammar
  match_word("GET "),
  match_before(" ",PathFile,_),
  match_word("HTTP/"),
  #>_version, % 
  !,
  split_path_file(PathFile,Ds,Fs),
  !,
  has_text_file_sufix(Fs,Suf),
  !,
  ( Suf=".pro"->http_process_local(Socket,Ds,Fs,Suf,Css)
  ; write('redirecting '),write_chars(Ds),write_chars(Fs),nl,
    http_send_line(Socket,"HTTP/1.0 302 Found"),
    make_cmd0(["Location: ",FallBackServer,Ds,Fs],Redirect),
    http_send_line(Socket,Redirect),
    http_send_line(Socket,"")
  ),
  close_socket(Socket).
\end{verbatim}}

Our HTTP server component fits in 76 lines of code, and can be used
to set up Prolog based Web processing by simply starting it in
a command window on any Unix machine or PC. This involves no execution overhead,
as is the case with CGI scripts. It basically offers the advantages
of Apache server side includes (SSIs) or Microsoft's active server pages (ASPs),
without requiring integration into a server, often subject to using the 
languages the server supports.

\subsection{Master Servers: Connecting a Web of Virtual Places}

A {\em virtual place} is implemented as a server listening
on a port which can spawn clients in the same
or separate threads interacting with other servers.


A master server on a `well-known' host/port
is used as a registration service to exchange identification information
among peers composed of clients and a server, 
usually running as threads of the same process.

As in the case of the HTTP server we can derive a master server by specializing
its interactor components through intuitionistic implications.

\section{Code migration and code acceleration techniques} \label{mobcode}

We have seen that setting up a self contained networking infrastructure
(Web protocols included) is a fairly simple task. The next step is  emphasizing mobile agent support, which is particularly promising in synergy with
knowledge processing capabilities  - a key strength of Logic Programming
systems.

\subsection{Lazy code fetching}

In BinProlog, code is fetched lazily over the network,
one predicate at a time, as needed by the execution flow.

Code is cached in a local database and then
dynamically recompiled on the fly if usage statistics
indicate that it is {\em not volatile} and it is {\em heavily used}
locally. 

The following operations

{\small \begin{verbatim}
host(Other_machine)=>>rload(File).
host(Other_machine)=>>code(File)=>>TopGoal.
\end{verbatim}}

\noindent allow fetching remote files {\tt rload/1} or
on-demand fetching of a predicate at a time from a remote
host during execution of {\tt TopGoal}. 

This is basically the same
mechanism as the one implemented for Java applet code fetching,
except that we have also implemented a caching mechanism,
at predicate level (predicates are cached as dynamic code
on the server to efficiently serve multiple clients).

\subsection{Dynamic recompilation}

Dynamic recompilation is used on the client side to speed-up heavily
used, relatively non-volatile predicates.  With dynamically recompiled
consulted code, listing of sources and dynamic modification to any
predicate is available, while the average performance stays close to
statically compiled code (usually within a factor of 2-3).

Our implementation of dynamic recompilation for BinProlog
is largely motivated by the difficulty/complexity of relying 
on the programmer
to specify execution methods for remote code.

The intuition behind the dynamic recompilation algorithm of BinProlog is
that {\em update} vs. {\em call} based {\em statistics} are associated
to each predicate declared or detected as dynamic.
Dynamic (re)compilation is triggered for relatively non-volatile predicates,
which are promoted on the {\em `speed-hierarchy'} to a faster
implementation method (interpreted \verb~->~ bytecode \verb~->~ native).
The process is restarted from the `easier to change' interpreted
representation, kept in memory in a compact form,
upon an update.

We can describe  BinProlog's dynamic {\em `recompilation
triggering statistics'} through a simple `thermostat' metaphor.
{\em Updates} (assert/retract) to a predicate have the effect of increasing its
associated `temperature', while {\em Calls}
will decrease it. Non-volatile (`cool') predicates 
are dynamically recompiled, while recompilation is avoided 
for volatile (`hot') predicates.
A {\em ratio} based on cooling factors (number of calls,
compiled/interpreted execution speed-up etc.) and
heating factors (recompilation time, number of updates etc.)
smoothly adjusts for optimal overall performance,
usually within a factor of 2 from static code.

\section{Computation Mobility with Threads and Continuations} \label{mobcomp}

\subsection{Why do computations need to be mobile?}

Advanced {\em mobile object} and {\em mobile agents} agent systems
have been built on top of Java's dynamic class loading and
its new reflection and remote method invocation classes.
IBM Japan's Aglets or General Magic's Odyssey
provide comprehensive mobility of code and data.
Moreover, data is encapsulated as state of objects.
This property allows us to protect its sensitive components
more easily. Distributed Oz/Mozart provides fully transparent
movement of objects over the network, giving the illusion
that the same program runs on all the computers.

So {\em why do we need} the apparently more powerful
concept of mobile ``live code'' i.e. mobile execution state?


Our answer to this question is that live mobile code is needed because
is still {\em semantically simpler} than
mobile object schemes. Basically, all that a programmer
needs to know is that his or her program has moved to
a new site and it is executing there. A unique (in our case
{\tt move\_thread}) primitive, with an
intuitive semantics, is all that needs to be learned.
When judging about how appropriate a language feature is,
we think that the way it looks to the end user is among the
most important ones. For this reason, mobile threads are
competitive with sophisticated {\em object mobility} constructs
on ``end-user ergonomy'' grounds,
while being fairly simple to implement, as we have shown,
in languages in which continuations can be easily
represented as data structures.

And {\em what if the host language does not offer first order
continuations}? A simple way around this is to implement
on top of it a script interpreter (e.g. a subset of Scheme or Prolog)
which does support them! As it is a good idea to limit code
migration to lightweight scripts anyway, this is a very
practical solution for either C/C++ or Java based mobile
code solutions, without requiring class-specific 
serialization mechanisms.

\subsection{Engines and Answer Threads} \label{engines}

Mobile computations really make sense only when multiple computation threads
can coexist at a given place. We build this infrastructure in two steps:
an engine encapsulating the state of an independent computation, and a thread
actually running it. Note that engines can also be used without multi-threading,
as a form of coroutining.

\subsubsection{Engines}
BinProlog allows launching
multiple Prolog engines having their own stack groups (heap, local stack
and trail). An engine can be seen as an abstract data-type which
produces a (possibly infinite) stream of solutions as needed.
To create a new engine, we use:

{\small \begin{verbatim}
  create_engine(+HeapSize,+StackSize,+TrailSize,-Handle)
\end{verbatim}}

\noindent or, by using default parameters for the stacks:

{\small \begin{verbatim}
  create_engine(-Handle)
\end{verbatim}}

\noindent The {\tt Handle} is a unique integer denoting the engine for further
processing.
To `fuel' the engine with a goal and an expected answer variable
we use:

{\small \begin{verbatim}
  load_engine(+Handle,+Goal,+AnswerVariable)
\end{verbatim}}

\noindent No processing, except the initialization of the
engine takes place, and no answer
is returned with this operation.

To get an answer from the engine we use:
{\small \begin{verbatim}
  ask_engine(+Handle,-Answer)
\end{verbatim}}

\noindent Each engine has its own heap garbage collection process
and backtracks independently using its choice-point stack and trail
during the computation of an answer.
Once computed, an answer is copied from an engine to its ``master''.

When the stream of answers reaches its end, \verb~ask_engine/2~
will simply fail. The resolution process in an engine
can be discarded at any time by simply loading another goal
with {\tt load\_engine/3}. This allows avoiding the
cost of backtracking, for instance in the case when a single
answer is needed, as well as garbage collection costs.

If for some reason we are not interested in the engine any more,
we can free the space allocated to the engine and completely discard it with:

{\small \begin{verbatim}
  destroy_engine(+Handle)
\end{verbatim}}

\noindent The following example \footnote{See more in files
library/engines.pl, progs/engtest.pl} in the BinProlog distribution \cite{bp7advanced}
shows a sequence of the previously described operations:

{\small \begin{verbatim}
 ?-create_engine(E),
   load_engine(E,append(As,Bs,[1,2]),As+Bs),
   ask_engine(E,R1),write(R1),nl,
   ask_engine(E,R2),write(R2),nl,
   destroy_engine(E).
\end{verbatim}}

\noindent Multiple  `orthogonal engines' as shown in Figure \ref{ortho}
enhance the expressiveness of
Prolog by allowing an AND-branch of an engine to
collect answers from multiple OR-branches of another engine.
They give to the programmer the means to see
as an abstract sequence and control, the answers
produced by an engine, in a way
similar to Java's {\tt Enumeration} interface.

\FIG{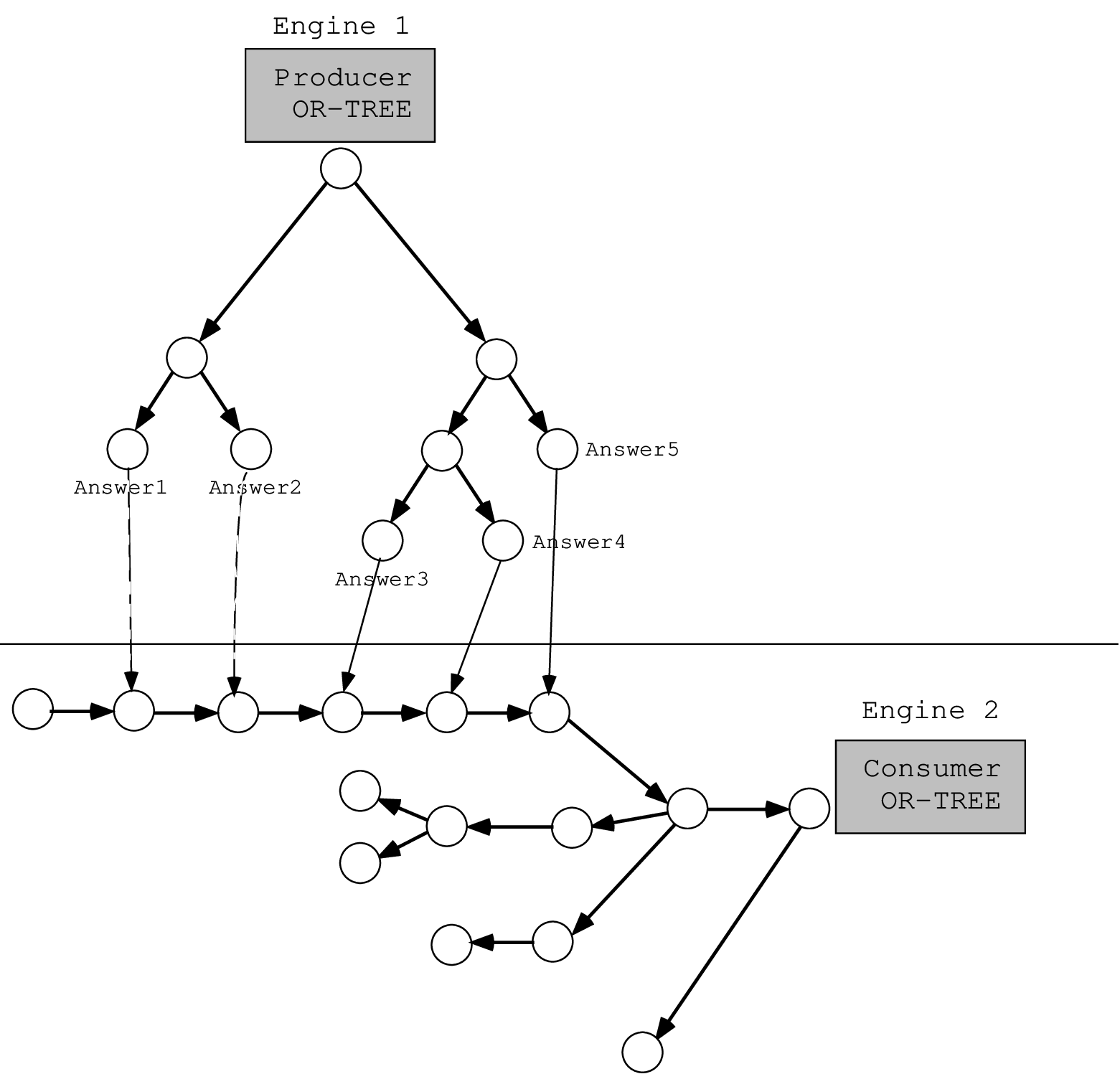}{2.75}{3}{Orthogonal Engines}

\subsubsection{Threads}

Engines can be assigned to their own thread by using
BinProlog's thread package (currently fully implemented on win32, Linux and Solaris platform).
A unique primitive is needed,
{\small \begin{verbatim}
   ask_thread(E,R)
\end{verbatim}}

\noindent
which launches a new thread {\tt R} to perform the computation of
an answer of engine {\tt E}.
On top of this facility each thread can implement a separate server or
client, or become the base of a mobile agent.

Thread synchronization is provided through monitor objects, handled
with:

{\small \begin{verbatim}
   synchronize_on(Monitor,Goal,Answer)
\end{verbatim}}

The thread waits until the monitor is free, executes Goal, frees the monitor and
returns Answer.

The thread attached to an engine can be obtained with: 

{\small \begin{verbatim}
   get_engine_thread(Engine,Thread)
\end{verbatim}}

and can be controlled directly with:

{\small \begin{verbatim}
  thread_suspend(Thread) % suspends the thread
  thread_resume(Thread)  % resumes a suspended thread
  thread_join(Thread)   % waits until a thread terminates
  thread_cancel(Thread) % discards a thread
\end{verbatim}}

\subsection{First order Continuations through Binarization} \label{bin}

Having first order continuations largely simplifies the implementation of
mobile code operations: a thread is suspended, its continuation is packed,
sent over the network and resumed at a different place.

We will shortly explain here BinProlog's continuation passing
preprocessing technique, which results in the availability of
continuations as data structures accessible to the programmer.

\paragraph{The binarization transformation}
Binary clauses have only one atom in the body
(except for some in-line `builtin' operations like arithmetics),
and therefore they need no `return' after a call.
A transformation introduced in \cite{Tarau90:PLILP} allows us to
faithfully represent logic programs with operationally equivalent
binary programs.

To keep things simple, we will describe our transformations in the case
of definite programs.
We will follow here the notations of \cite{pt93b}.

Let us define the {\em composition} operator \mf{\oplus} 
that combines clauses by unfolding the leftmost body-goal
of the first argument.

Let {\tt A$_0$:-A$_1$,A$_2$,\ldots,A$_n$} and 
{\tt B$_0$:-B$_1$,\ldots,B$_m$} be two clauses (suppose $n>0, m\ge 0$). We define 
\begin{flushleft} {\tt (A$_0$:-A$_1$,A$_2$,\ldots,A$_n$)} \mf{\oplus} 
{\tt (B$_0$:-B$_1$,\ldots,B$_m$) = (A$_0$:-B$_1$,\ldots,B$_m$,A$_2$,\ldots,A$_n$)}$\theta$
\end{flushleft}
with $\theta$ = mgu({\tt A$_1$},{\tt B$_0$}). If the atoms {\tt A$_1$} and
{\tt B$_0$} do not unify, the result of the composition is denoted as \mf{\bot}.
Furthermore, as usual, we consider {\tt A$_0$:-true,A$_2$,\ldots,A$_n$} 
to be equivalent to {\tt A$_0$:-A$_2$,\ldots,A$_n$}, and for any clause {\tt C}, {\tt $\bot$ $\oplus$ C = C $\oplus$ $\bot$ = $\bot$}.
We assume that at least one operand has been renamed to a variant with
 variables standardized apart. 

This Prolog-like inference rule is called LD-resolution. It has
the advantage of giving a more accurate description of Prolog's operational semantics than SLD-resolution.
Before introducing the binarization transformation, we describe two
auxiliary transformations.

The first transformation converts facts into rules by  giving
them the atom {\tt true} as body. E.g., the fact {\tt p} is
transformed into the rule {\tt p :- true}.

The second transformation, inspired by \cite{Warren82},
eliminates the metavariables by wrapping them in a {\tt call/1} goal.
E.g., the rule {\tt and(X,Y):-X, Y} is transformed into {\tt 
and(X,Y) :- call(X), call(Y).}

The transformation of \cite{Tarau90:PLILP}
({\em binarization}) adds continuations
as  extra   arguments  of   atoms  in a way that  preserves
also first argument indexing.

Let   P be  a definite  program  and \mf{Cont}  a  new
variable. Let  \mf{T} and \mf{E=p(T_1,...,T_n)} be  two 
expressions.\footnote{Atom or term.} We  denote by
\mf{\psi(E,T)} the expression \mf{p(T_1,...,T_n,T)}. 
Starting with the clause

\verb~(C)~  \qquad \mf{A :- B_1,B_2,...,B_n.}

\noindent we construct the clause

\verb~(C')~  \qquad \mf{\psi(A,Cont) :-
 \psi(B_1,\psi(B_2,...,\psi(B_n,Cont))).}
                         
\noindent
The set $P'$ of all clauses \verb~C'~ obtained from the clauses of P is called
the binarization of P. 

The following example shows the result of this
transformation on the well-known `naive reverse' program:

{\small \begin{verbatim}
   app([],Ys,Ys,Cont):-true(Cont).
   app([A|Xs],Ys,[A|Zs],Cont):-
     app(Xs,Ys,Zs,Cont).

   nrev([],[],Cont):-true(Cont).
   nrev([X|Xs],Zs,Cont):-
     nrev(Xs,Ys,app(Ys,[X],Zs,Cont)).
\end{verbatim}}

{\flushleft The} transformation preserves a
strong operational equivalence with the
original program with respect to the LD resolution rule, which
is {\em reified} in the syntactical structure of the
resulting program, i.e. each resolution step
of an LD derivation on a definite program $P$
can be mapped to an SLD-resolution step of the binarized program $P'$.

Clearly, continuations become explicit in the binary version of the program.
We have devised a technique to access and manipulate them in an intuitive
way, by modifying BinProlog's binarization preprocessor.
Basically, the clauses constructed with {\tt ::-} instead of {\tt :-}
are considered as being already in binary form, and not subject
therefore to further binarization. By explicitly accessing
their arguments, a programmer is able to access and modify the
current continuation as a `first order object'.
Note however that code {\em referring} to the continuation
is also {\em part} of it, so that some care should be taken in
manipulating the circular term representing the continuation
from `inside'.

\subsection{Mobile threads: Take the {\em Future} and Run} \label{mobthreads}

As continuations (describing {\em future} computations
to be performed at a given point)
are first order objects in BinProlog,
it is easy to extract from them a conjunction of goals
representing
{\em future} computations intended to be performed at
another site,
send it over the network and resume working on it
at that site.
The natural unit of mobility is a {\em thread}
moving to a server executing multiple
local and remotely originated threads.
{\em Threads communicate with their local and remote
counterparts, listening on ports
through the Linda protocol, as described in \cite{dbt95a}}.
This combination of Linda based coordination and thread
mobility is intended to make building complex, pattern based
agent scripts fairly easy.

\subsubsection{Capturing continuations} \label{capt}

Before moving to another site, the current continuation
needs to be captured in a data structure (see Appendix I).
For flexibility, a wrapper {\tt capture\_cont\_for/1}
is used first to
restrict the scope of the continuation to
a (deterministic) toplevel {\tt Goal}. This avoids taking
irrelevant parts of the continuation (like prompting the user
for the next query) to the remote site inadvertently.

A unique logical variable is used through a backtrackable linear
assumption {\tt cont\_marker(End)} to mark the end
of the scope of the continuation with {\tt end\_cont(End)}.

>From inside the continuation, {\tt call\_with\_cont/1} is used to
extract the relevant segment of the continuation.  Towards
this end, {\tt consume\_cont(Closure,Marker)} extracts a conjunction of
goals from the current continuation until Marker is reached, and then it
applies {\tt Closure} to this conjunction (calls it with the
conjunction passed to {\tt Closure} as an argument).

Extracting the continuation itself
is easy, by using BinProlog's ability to
accept user defined binarized clauses
(introduced with ::- instead of :-),
that can access the continuation as a `first order' object:

{\small \begin{verbatim}
  get_cont(Cont,Cont)::-true(Cont).
\end{verbatim}}

\subsubsection{The Continuation Moving Protocol} \label{contmoving}

Our continuation moving protocol can be described easily in terms
of synchronized {\em source side}\footnote{which will be
also shortly called the {\em base} of the mobile thread},
and {\em target side} operations.

\paragraph{Source side operations}
\begin{itemize}
\item wrap a Goal
with a unique terminator marking the end of the continuation to be
captured, and call it with the current continuation available
to it through a linearly assumed fact\footnote{BinProlog's linear assumptions
 are backtrackable additions to the database, usable at most once.
}

\item reserve a free port P for the future code server
\item schedule on the target server a sequence of actions
 which will lead to resuming the execution from right after the
 {\tt move\_thread} operation (see target
side operations), return and become a code server allowing the mobile
thread to fetch required predicates one a time
\end{itemize}
\paragraph{Target side operations} are scheduled as a sequence of goals
extracted from the current continuation at the {\em source side }, and received
over the network together with a small set of synchronization commands:

\begin{itemize}
\item schedule as delayed task a sequence of goals received from the
source side and return
\item wait until the {\em source side} is in server mode
\item set up the back links to the source side as assumptions
\item execute the delayed operations representing the moved continuation
\item fetch code from the source side as needed for execution of the goals
of the moved continuations and their subcalls
\item shut down the code server on the source side
\end{itemize}

Communication between the base and the target side
proceeds through {\em remote predicate calls}
protected with {\em dynamically generated passwords}
shared between the two sides before the migratory
component ``takes off''.

Initially the target side waits in server mode. Once the continuation
is received on the target side, the source side switches to server mode
(unless it already contains a running server thread),
ready to execute code fetching and persistent database
update requests from its mobile counterpart on the target side.

Fig. \ref{mob} shows the connections between a mobile thread
and its base.

Note that when the base turns into a server, it offers its
{\em own code} for remote use by the moved thread - a kind of
virtual ``on demand'' process cloning operation, one step
at a time. Since the server actually acts as a code cache,
multiple moving threads can benefit from this operation.
Note also that only predicates needed for the migratory
segment of the continuation are fetched. This ensures
that migratory code is kept lightweight for most mobile
applications. Synchronized communication, using Linda
operations, can occur between the mobile thread and
its base server, and through the server, between
multiple mobile threads which have migrated to various
places. Note that {\em servants} described in section \ref{servant}
can also be used to {\em emulate servers} in case of the
ptocess is hosted in a component unable to listen
on a server port.

As our networking infrastructure, our {\em mobile threads} are
platform independent. Like Java, BinProlog \cite{bp7user} is a platform
independent emulator based language.
As a consequence, a thread can start on a Unix machine and
move transparently to a Windows NT system and back.
For faster, platform specific
execution, BinProlog provides compilation to C of static code using
an original partial translation technique described in \cite{tdb95rev}.

\FIG{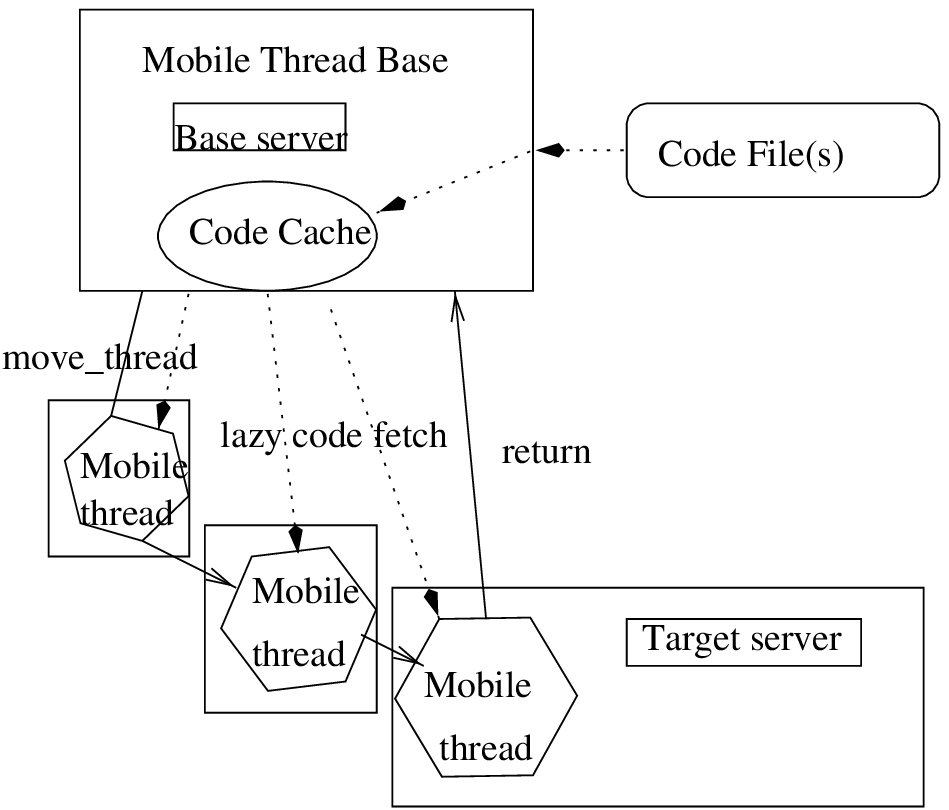}{2}{2.5}{Launching a mobile thread from its base}

\subsubsection{Emulating computation mobility through control mobility} \label{mobemu}

As shown in \cite{TDB:97},
part of the functionality of {\em mobile computations} can be emulated
in terms of remote predicate calls combined with remote
code fetching.
An implicit {\em virtual place} (host+port) can be set as the target
of the remote calls.
Then, it is enough to send the top-level goal
to the remote side and have it fetch the code as needed from
a server at the site from where the code originates.

Note however that this is less efficient in terms of network
transactions and less reliable than
sending the full continuation at once as with
our {\em mobile threads}.


\subsection{Mobile Agents} \label{mobags}

{\em Mobile agents} can be seen as a collection
of synchronized {\em mobile threads} sharing common
state \cite{TD96:coord}. Agents execute a set of goals, 
possibly spread over a set of different processes.
In addition to usual Prolog computations, 
agents perform remote and local transactions in coordination with other agents,
and react to event driven changes occurring on blackboards. 

Mobile agents are implemented by sending out {\em mobility threads} 
to a set of servers\footnote{possibly filtered down
to a relevant subset using a `channel'-like pattern} 
registered to a given master server.
A pyramidal deployment strategy can be used
to efficiently implement, for instance,
{\em push technology} through mobile
agents. Inter-agent communication can be achieved either by rendez-vous
of two mobile threads at a given site, by communicating through
a local Prolog database, or through the base server known to
all the deployed agents.
Communication with the base server
is easily achieved through remote predicate calls with
{\tt remote\_run}.
Inter-agent communication can be achieved either by rendez-vous
of two mobile threads at a given site, by communicating through
a local blackboard, or through the base server known to
all the deployed agents.
Basic security of mobile agents is achieved with
randomly generated passwords,
required for {\tt remote\_run/1} operations, and
by running them on a restricted BinProlog machine,
without user-level file write and
external process spawn operations.
For more details on recent developments of
mobile agent infrastructures and their applications, extending
the framework presented in this paper, we refer 
to \cite{tarau:paam99,tarau:shaker,mikler:hpc99,tarau:dipcl99,RDT99}.

In the latest version of BinProlog \cite{bp7advanced} 
two simple {\bf move/0} and {\bf return/0} 
operations can be used to transport a mobile agent thread to a server 
(listening on a default host/port) and back. The anaphoric {\em there}
indicates that the mobile thread should be created on the server. Alternatively,
{\em here} would indicate execution on a local thread.
In this mobility scheme, the client simply waits until computation 
completes, when bindings for 
the first solution are propagated back.

{\small \begin{verbatim}
Window 1: a mobile thread 

?-there,move,println(on_server),member(X,[1,2,3]),
        return,println(back). 
back
X=1

Window 2: a server 

?-trust. 
on_server 
\end{verbatim}}

In case return is absent, computation proceeds to the 
end of the transported continuation.

Note that mobile computation is more expressive and more 
efficient than remote predicate calls as such. 
Basically, it {\em moves once}, 
and executes on the server {\em all future computations} of the current 
AND branch until a return instruction is hit,
when it takes the remaining continuation and comes back. 
This can be seen by comparing real time execution speed for: 

{\small \begin{verbatim}
?-for(I,1,1000),remote_run(println(I)),fail. 

?-there,move,for(I,1,1000),println(I),fail. 
\end{verbatim}}
 
While the first query uses {\tt remote\_run/1} each time to send a remote task to the server,
the second moves once the full computation to the server where it executes without
further requiring network communications.
Note that the {\tt move/0, return/0} pair cut nondeterminism for the transported segment
of the current continuation. This avoids having to transport state of the choice-point stack
as well as implementation complexity of multiple answer returns and tedious 
distributed backtracking synchronization. Surprisingly, this is not a strong limitation,
as the programmer can simply use something like:

{\small \begin{verbatim}
?-there,move,findall(X,for(I,1,1000),Xs),return,member(X,Xs).
\end{verbatim}}

\noindent to emulate (finite!) nondeterministic remote execution, 
by collecting all solutions
at the remote side and exploring them through (much more efficient) local
backtracking after returning.

\section{Related work} \label{rel}

Remote execution and code migration techniques are
pioneered by \cite{Alme85,Jul88,Stam90}. Support for
remote procedure calls (RPC) are part of major operating
systems like Sun's Solaris and Microsoft's Windows NT.

A very large number of research projects
have recently started on mobile computations/mobile agent programming.
Among the pioneers, Kahn and Cerf's Knowbots \cite{knowbots}.
Among the most promising recent developments,
Luca Cardelli's Oblique project at  Digital and
mobile agent applications \cite{migratory}
and IBM Japan's aglets \cite{aglets}.
Mobile code technologies are pioneered by General Magic's
Telescript (see \cite{odissey} for their last Java based
{\em mobile agent} product). General Magic's Portico
software combines mobile code technologies and
voice recognition based command language (MagicTalk).

Another mobility framework, sharing some of our objectives
towards transparent high level distributed programming,
is built on top of Distributed Oz \cite{DOZmobs,distoz97}
a multi-paradigm language, also including a logic programming
component.
Although thread mobility is not implemented in Distributed Oz 2,
some of this functionality can be emulated in terms of
network transparent mobile objects.
The illusion of a unique application which transparently
runs on multiple sites makes implementing shared
multi-user applications particularly easy.
We can emulate this by implementing mobile agents (e.g.
avatars) as mobile threads with parts of the shared world {\em visible}
to an agent represented as dynamic facts, lazily replicated through our
lazy code fetching scheme when the agent moves.
Both Distributed Oz 2 and our BinProlog based infrastructure
need a full language processor
(Oz 2 or BinProlog) to be deployed at each node.
However, assuming that a Java processor is already installed,
our framework's Java client (see \cite{TDB:97,TDBwetice:97})
allows this functionality to be available through
applets attached to a server side BinProlog thread.

Implementation technologies for mobile code are studied in 
\cite{adl-tabatabai96:efficient}.
Early work on the Linda coordination framework \cite{linda89,CasCia96,sharedpro91} has
shown its potential for coordination of multi-agent systems.
The logical modeling and planning aspects of computational Multi-Agent systems have been pioneered
by \cite{cohen:79a,cohen:89a,kowalski:91a,wooldridge:92a,cohen:94a,cohen:95a,lesperance:95a,chaibdraa:94a}.
A survey of Logic Programming approaches to Web applications in terms of a
classification into client-based systems, 
server-side systems, and peer-to-peer systems is provided in \cite{loke98}. 


Let us point out a key difference between our framework and a typical
Java based mobile code system.
Conventional mobile code systems like IBM's Aglets \cite{aglets} require 
serialization hints from the programmer and do not implement a fully generic
reflective computation mobility infrastructure. Aglets do not provide code
mobility as they assume that code is already available at the destination site.
In practice this means that the mobile code/mobile computation layer is
not really transparent to the programmer. 

In contrast, our architecture is based on building an autonomous layer consisting of
a reflective interpreter which provides the equivalent of implicit serialization 
and supports orthogonal transport mechanisms for
data, code and computation state. 
The key idea is simply that by 
introducing interpreters spawned as threads by a server at each networked site, 
{\em computation mobility} 
at object-level is mapped to {\em data mobility} at meta-level
in a very straightforward way. 
A nice consequence is transport independence
coming from the unifrom representation of data, code and computation state
allowing Corba, RMI, HLA, as well as plain or multicast sockets
to be used interchangeably as our transport mechanism.

\section{Conclusion} \label{conc}

We have described how mobile threads are implemented by capturing
first order continuations in a data structure sent over the network.
Supported by {\em lazy code fetching} and
{\em dynamic recompilation},
they have been shown to be
an effective framework for
implementing mobile agents.

The techniques presented here are not (Bin)Prolog specific.
The most obvious porting target of our design is to functional
languages featuring first order continuations and threads.
Another porting target is Java and similar object oriented languages
having threads, reflection classes and
remote method invocation.
We are working on a Java based component
using an embedded continuation passing Prolog interpreter
which is already able to interoperate with BinProlog 
(latest version at \url{http://www.binnetcorp.com/Jinni} ).
An interesting application is using BinProlog as an accelerator
for Java based threads through migration to BinProlog, execution of
a computationally intensive task and return to the Java component.

\subsection*{Acknowledgement} We thank the anonymous referees
for their valuable suggestions and comments. We thank
for support from NSERC (grants OGP0107411 and 611024).




\bibliography{tarau,biblio,mobile,agents}

\section*{Appendix I: Capturing First Order Continuations in BinProlog}
{\small \begin{verbatim}
% calls Goal with current continuation available to its inner calls
capture_cont_for(Goal):-
  assumeal(cont_marker(End)),
    Goal,
  end_cont(End).

% passes Closure to be called on accumulated continuation
call_with_cont(Closure):-
  assumed(cont_marker(End)),
  consume_cont(Closure,End).
  
% gathers in conjunction goals from the current continuation
% until Marker is reached when it calls Closure on it
consume_cont(Closure,Marker):-
  get_cont(Cont),
  consume_cont1(Marker,(_,_,_,Cs),Cont,NewCont), % first _
  call(Closure,Cs),                              % second _
  % sets current continuation to leftover NewCont    
  call_cont(NewCont).                            % third _

% gathers goals in Gs until Marker is hit in continuation Cont
% when leftover LastCont continuation (stripped of Gs) is returned
consume_cont1(Marker,Gs,Cont,LastCont):-
   strip_cont(Cont,Goal,NextCont),
   ( NextCont==true-> !,errmes(in_consume_cont,expected_marker(Marker))
   ; arg(1,NextCont,X),Marker==X->
     Gs=Goal,arg(2,NextCont,LastCont)
   ; Gs=(Goal,OtherGs),
     consume_cont1(Marker,OtherGs,NextCont,LastCont)
   ).

% this `binarized clause' gets the current continuation
get_cont(Cont,Cont)::-true(Cont).

% sets calls NewCont as continuation to be called next
call_cont(NewCont,_) ::- true(NewCont).
\end{verbatim}}

\newpage
\section*{Appendix II: Thread Mobility in BinProlog}
{\small \begin{verbatim}
% wraps continuation of current thread to be taken
% by inner move_thread goal to be executed remotely 
wrap_thread(Goal):-
  capture_cont_for(Goal).

% picks up wrapped continuation,
% jumps to default remote site and runs it there
move_thread:-
  call_with_cont(move_with_cont).
  
% moves to remote site goals Gs in current continuation
move_with_cont(Gs):-
  % gets info about this host
  detect_host(BackHost),
  get_free_port(BackPort),
  default_password(BackPasswd),
  default_code(BackCode),
  % runs delayed remote command (assumes is with +/1)
  remote_run(
     +todo(
       host(BackHost)=>>port(BackPort)=>>code(BackCode)=>>(
         sleep(5), % waits until server on BackPort is up
         % runs foals Gs picked up from current continuation 
         (Gs->true;true), % ignores failure
         % stops server back on site of origin
         stop_server(BackPasswd)
       )
     )
  ),
  % becomes data and code server for mobile code until is
  % stopped by mobile code possessing password
  server_port(BackPort)=>>run_unrestricted_server.
\end{verbatim}}
\end{document}